\documentstyle[amstex,amssymb,12pt,fleqn]{article}

\topmargin - 0.8cm

\renewcommand{\theequation}{\arabic{section}.\arabic{equation}}
\renewcommand{\thesection}{\arabic{section}.}

\catcode`@=11 \@addtoreset{equation}{section} \catcode`@=12
\mathchardef\SGamma="7100

\begin{document}

\title{{\bf New Nonlocal Effective Action}}
\author{A.O. Barvinsky$^{1}$ and V.F.Mukhanov$^{2}$ \\
$^{1}$Theory Department, Lebedev Physics Institute,\\
Leninsky prospect 53, Moscow 117924;\\
e-mail: barvin@td.lpi.ac.ru\\
$^{2}$Sektion Physik, LMU, Theresienstr.37,Munich,Germany; \\
e-mail: mukhanov@theorie.physik.uni-muenchen.de}
\maketitle

\begin{abstract}
We suggest a new method for the calculation of the nonlocal part of the
effective action. It is based on resummation of perturbation series
for the heat kernel and its functional trace at large values of the proper
time parameter. We derive a new, essentially nonperturbative, nonlocal
contribution to the effective action in spacetimes with dimensions $d>2$.
\end{abstract}

\section{Introduction}

Effective action is among the fundamental ideas of modern quantum
field theory. Calculated analytically for a given background
field, it gives information about the induced energy-momentum
tensor of quantum fields and quantum corrections to the classical
equations of motion. The nonlocal part of the effective action
should contain, for instance, particle creation effects. For black
holes it should be able to account simultaneously for vacuum
polarization and asymptotic Hawking radiation. Various important
applications of the effective action can also be found in
fundamental string theory. The Lorentzian effective action, which
we actually need, can be obtained from the Euclidian one,
$\SGamma[\,\phi\,]$, via analytic continuation. In turn,
$\SGamma[\,\phi\,]$ can be defined by the following path integral
    \begin{equation}
    \exp\Big(-\SGamma[\,\phi(x)\,]\Big)= \int D\varphi\,
    \exp \left(-S[\,\varphi\,]
    +(\varphi -\phi)\frac{\delta\SGamma[\,\phi\,]}{\delta\phi}
    \,\right) ,  \label{efacdef}
    \end{equation}
where $\phi(x)$ is a given mean field, and the functional
integration over quantum fields $\varphi(x)$ is assumed. The
general semiclassical expansion of $\SGamma[\,\phi\,]$ begins with
the one-loop contribution, which is given by the gaussian path
integral
    \begin{equation}
    \exp(-\SGamma[\,\phi(x)\,])=
    \int D\varphi \exp \left( -\frac{1}{2}\int dx\,
    \sqrt{g}\, \varphi(x)\,\widehat{F}
    \left(\nabla,\phi(x)\right)\varphi(x)\,\right)  \label{oneloop}
    \end{equation}
The operator $\widehat{F}(\nabla,\phi(x))$ here determines the
propagation of small field disturbances $\varphi(x)$ on the
background of $\phi(x)$ and in bosonic case can generically be
written down as
    \begin{equation}
    \widehat{F}=-\Box +V(x),  \label{opF}
    \end{equation}
where $\Box =$ $\nabla ^{2}\equiv g^{\mu \nu }\nabla _{\mu }\nabla
_{\nu }$ is the Laplacian in the Euclidean field theory, which
becomes the d'Alembertian when analytically continued to the
Lorentzian sector, and $V(x)$ is the potential term. Note that,
for some fields, the one-loop contribution is exact, for instance,
for the scalar field without self-coupling. For a properly defined
measure the gaussian integral (\ref{oneloop}) can be formally
calculated as
    \begin{equation}
    \SGamma=\frac{1}{2}\ln \left( \prod_\lambda \lambda\right)
    =\frac{1}{2}\sum_\lambda \ln
    \lambda =\frac12\,{\rm Tr}\,\ln \widehat{F},  \label{act1}
    \end{equation}
where $\lambda$ are the eigenvalues of the operator $\widehat{F}$
corresponding to appropriately normalized eigenfunctions
$\phi_\lambda(x)$, $\int dx\,\sqrt{g}\,\phi _\lambda(x)
\phi_{\lambda'}(x)=\delta _{\lambda\lambda'}$. Here the functional
trace ${\rm Tr}$ does not depend on a particular basis in the
functional space of disturbances $\varphi$ and, therefore, in an
appropriate representation it reduces to the integral over spatial
coordinates $x$ of the diagonal element of the operator kernel.

The effective action (\ref{act1}) is, of course, ultraviolet
divergent and should be regularized, with the subsequent
interpretation of explicitly isolated divergences in terms of
infinite renormalizations of the coupling constants of the theory.
These divergences are well understood and it is unlikely that
anything new can be added here. Therefore, we concentrate on more
interesting finite, and generally nonlocal, contributions to
one-loop effective action. These contributions depend on infrared
properties of the theory and contain nontrivial information about
real physical effects. Analytical calculational schemes for
$\SGamma$ are usually based on the following integral
representation of the functional trace of $\widehat{F}$
    \begin{equation}
    {\rm Tr}\,\ln \widehat{F}=
    -\int\limits_{0}^{\infty }\frac{ds}{s}\,
    {\rm Tr}\,e^{-s\widehat{F}},  \label{intrepr}
    \end{equation}
where all local divergences can be easily isolated with the aid of
dimensional regularization. The kernel
    \begin{equation}
    K(s|\,x,y)\equiv
    \exp(-s\widehat{F})\,\delta ^{(d)}(x,y), \label{kern}
    \end{equation}
where $d$ is spacetime dimensionality, obviously satisfies the
heat kernel equation
    \begin{equation}
    \frac{\partial }{\partial s}K(s|\,x,y)=
    -\widehat{F}K(s|\,x,y)\equiv (\Box
    -V(x))K(s|\,x,y),  \label{heateq}
    \end{equation}
with the initial condition
    \begin{equation}
    K(0|\,x,y)=\delta ^{(d)}(x,y)  \label{incond}
    \end{equation}%
at $s=0.$ The auxiliary parameter $s$ is usually called the proper
time. Thus, calculating the effective action can be reduced to the
solution of the Cauchy problem for $K(s|\,x,y)$. In fact, what we
actually need is the coincidence limit of this function, since in
the representation we used, the functional trace of the operator
$e^{-s\widehat{F}}$ corresponds to the integration of the diagonal
elements of $K$ over the spacetime coordinates $x$, so that
    \begin{equation}
    \SGamma=-\frac{1}{2}\int dx\,
    \left( \int_{0}^{\infty }\frac{ds}{s}
    K(s|\,x,x)\right).                     \label{act2}
    \end{equation}

It is clear that the success of the calculation of mainly depends
on our ability to find an analytical solution of the heat kernel
equation and carry out the integration over the proper time in
(\ref{act2}). The integral is obviously divergent as $s
\rightarrow 0$. As we have already mentioned above, this
divergence can be easily isolated and interpreted in terms of the
local ultraviolet properties of the theory. On the other hand, the
behavior of the integral at infinity, $s\rightarrow\infty$,
determines infrared properties of the theory and carries the
physical information, e.g., particle creation. If the field has a
big positive mass the proper time integral is convergent at
$s\rightarrow \infty$. However, in the case of massless fields,
the situation is much less trivial. The infrared convergence here
depends on the approximation scheme used to calculate
$K(s|\,x,y)$. Then, it is often even unclear to what extent the
obtained effective action reflects the physical properties of the
theory rather than the features of the approximation scheme used.

Below, we discuss the known calculational techniques, namely, the
local Schwinger-DeWitt expansion \cite{DeWitt,PhysRep}, nonlocal
covariant perturbation theory \cite{CPTI,CPTII,CPTIII} and the
modified gradient expansion \cite{MWZ} and point out why all of
them fail when applied to interesting physical problems. Instead
of them, we suggest a new method based on resummation of
perturbation series and calculate new, essentially nonperturbative
terms in the effective action. This method becomes indispensable
in low-dimensional models $(d\leq 2)$ where all previously known
techniques are inapplicable. In this paper, we demonstrate how our
method works in flat space of dimension $d>2,$ while the
generalization to the curved space and low-dimensional case will
be considered in \cite{infrII}.

One of the main results of this paper is an exact (nonperturbative
in $V$) late-time asymptotics for the heat kernel which in a
spacetime of dimension $d>2$ for the coincidence limit takes the
following form
    \begin{equation}
    K(s|\,x,x)=\frac{1}{(4\pi s)^{d/2}}
    \left( 1+\frac{1}{\Box -V}V(x)\right)^{2},
    \,\,\,s\rightarrow \infty .           \label{ass1a}
    \end{equation}
This asymptotics can also be found when the arguments of
$K(s|\,x,y)$ are different - see Sect.4 for details. Another
important result is the nonperturbative in potential expression
\begin{equation}
{\rm Tr}\,K(s)=\frac{1}{(4\pi s)^{d/2}}\int dx\,
\left\{ 1-s\Big(V(x)+V\frac{1%
}{\Box -V}V(x)\Big)\right\} \,\,\,s\rightarrow \infty
\label{latetr}
\end{equation}
for the functional trace of $K(s)$\footnote{Note that this
expression for ${\rm Tr}\,K(s)$ cannot be obtained directly by
integrating the asymptotics (\ref{ass1a}) over the whole spacetime
because for a given $s$ this asymptotics fails at
$\left|x\right|^{2}>s$. Its derivation is given in Sect. 4 and
Appendix B, where we show that the expressions (\ref{ass1a}) and
(\ref{latetr}) are in complete agreement with each other.}.

To avoid excessive use of integration signs we employ here and
throughout the paper the following shorthand notation
\begin{equation}
\frac{1}{\Box -V}\,J(x)\equiv \int dy\,G(x,y)\,J(y),  \label{notation}
\end{equation}%
where $G(x,y)$ is the Green's function of the operator
$-\widehat{F}=\Box-V$ with zero boundary conditions at spacetime
infinity, that is
\begin{equation}
(\Box-V)\,G(x,y)=\delta ^{(d)}(x,y),\,\,\,G(x,y)\rightarrow
0,\,\,\,|x|\rightarrow \infty ,  \label{GR}
\end{equation}%
and $J(x)$ can be any function of various field quantities like
powers of potential, its derivatives, etc. We always presume that
the spacetime has the positive definite (Euclidean) signature, so
that the Laplacian $\Box $ is negative definite assuming zero
boundary conditions at infinity. Moreover we consider only
non-negative potentials $V\left( x\right) \geq 0,$ so that the
whole operator $\widehat{F}=-\Box +V$ is positive definite.
Therefore, the Green's function (\ref{GR}) is uniquely defined and
guarantees that the nonlocal expression (\ref{notation}) makes
sense for $d>2$.

As we shall see, the asymptotics (\ref{ass1a}) and (\ref{latetr})
are the corner stone of the technique we develop for the
calculation of nonlocal contributions to the effective action. In
particular, they lead to essentially nonperturbative terms which
can be explicitly calculated for two broad classes of potentials
with compact support, namely, for those ones which are,
respectively, very small or very big in units of the inverse size
of their support. For small potentials we get the terms which {\em
replace} the conventional Coleman-Weinberg contribution to the
effective action.  In four dimensions, for instance, these terms
read
    \begin{equation}
    \Delta \SGamma\!=\!\frac{1}{64\pi ^{2}}\!\int d^4x\,V^{2}
    \ln \Big(\int d^4y\,V^{2}\Big)\!
    -\!\frac{1}{64\pi ^{2}}\!\int d^4x\,V^{2}
    \ln \Big(\int d^4y\,
    V\frac{\mu ^{2}}{V-\Box}V\Big),  \label{4deff}
    \end{equation}
where the mass parameter $\mu ^{2}$ reflects the usual ultraviolet
renormalization ambiguity. On the contrary, in the case of big
potentials the Coleman-Weinberg action is {\em supplemented} by
the nonlocal term of the form
    \begin{equation}
    \Delta \SGamma=\frac{1}{64\pi ^{2}}
    \int\limits_{|x|\leq R}\!d^4x\,
    \Big<V+V\frac{1}{\Box -V}V\Big>^{2}.
    \end{equation}
Here $R$ is the size of the compact support of $V(x)$, that is, $V(x)=0$ at $%
|x|>R$, and $\big<...\big>$ denotes the spacetime averaging of the
corresponding quantity over this compact domain. The obtained
expressions are both nonlocal and non-analytic in the potential
$V(x)$.

The paper is organized as follows. In Sect.2 we consider the known
approximation schemes and discuss their applicability in the
infrared region. Sect.3 is devoted to the nonlocal and nonlinear
resummation of the Schwinger-DeWitt perturbation series,
corresponding to the so-called connected graph expansion of the
heat kernel. In Sect.4, as an extention of this resummation, we
derive the asymptotics (\ref{ass1a}) and discuss its relation to
the functional trace (\ref{latetr}) of the heat kernel. The
nonperturbative, nonlocal contributions to the effective action
are obtained in Sect.5 with the aid of the new technique based on
a piecewise smooth approximation for the heat kernel. In two
appendices we give details of the resummation technique and derive
from the covariant perturbation theory the asymptotics of the heat
kernel trace (\ref{latetr}) up to the first subleading order in
$1/s$ inclusive.

\section{Approximation schemes and infrared properties of the effective
action}

In flat space, which we consider in this paper, the solution of
the heat kernel equation can be easily found if the potential
vanishes. For an arbitrary spatially dependent potential the
analytical expressions are, of course, available only in certain
approximations. In the general case, it is convenient to factorize
the ''zero potential'' part of the solution explicitly and use the
following ansatz for $K(s|\,x,y)$
    \begin{equation}
    K(s|\,x,y)=\frac{1}{\left( 4\pi s\right)^{d/2}}
    \exp \left[ -\frac{\left|x-y\right|^{2}}{4s}\right]
    \Omega (s|\,x,y),                           \label{ansatz}
    \end{equation}
where the factor singular in $s$ guarantees that the initial
condition (\ref{incond}) is satisfied, provided that $\Omega$ is
analytic in $s$ at $s=0$ and $\Omega (0|\,x,y)=1$. If $V=0$, then
$\Omega \equiv 1$ and hence all nontrivial information about the
potential is encoded in the deviation of $\Omega$ from unity.

The most well-known approximation used for the calculation of
$K(s|\,x,y)$ is the so-called local Schwinger-DeWitt expansion,
where $\Omega$ is written down as a series in growing powers of
the proper time $s$. This expansion is a very powerful tool for
revealing local ultraviolet properties of the theory. However,
when applied in the infrared region, it gives a finite result only
for massive fields. If the potential $V(x)$ has a large positive
constant part, that is,
\begin{equation}
V(x)=m^{2}+v(x),  \label{pot1}
\end{equation}
where $m^{2}$ is the squared mass of the field, then the function
$\Omega(s|\,x,y)$ contains an overall exponential factor
$e^{-sm^{2}}$ and the independent of mass part is expanded in
powers of $s$
    \begin{equation}
    \Omega (s|\,x,y)=e^{-sm^{2}}
    \sum\limits_{n=0}^{\infty }a_{n}(x,y)\,s^{n}.    \label{dewittexp}
    \end{equation}
Here $a_{n}(x,y)$ are the two-point Schwinger-DeWitt coefficients,
whoes coincidence limits $\left( x\rightarrow y\right)$ are
explicitly calculable in general field theories, including
gravity. Substituting (\ref{dewittexp}) in (\ref{ansatz}) and then
the obtained expression at $x=y$ in (\ref{act2}) one gets:
    \begin{equation}
    \SGamma=-\frac{1}{2\left( 4\pi \right) ^{d/2}}\int
    dx\sum\limits_{n=0}^{\infty }\,\left( \int_{0}^{\infty}
    s^{n-d/2-1}e^{-sm^{2}}ds\right) a_{n}(x,x).       \label{dewittact1}
    \end{equation}

It is important that the exponent $e^{-sm^{2}}$ is not expanded
here in powers of $s$. Therefore, in the proper time integral it
provides a cutoff at the upper limit, so that the powers of $s$ in
this expansion get effectively replaced by powers of $1/m^{2}$.
The first $(d/2+1)$ integrals in (\ref{dewittact1}) diverge at
$s\rightarrow 0$ and should be regularized. To do that, we apply
the dimensional regularization method. Namely, by replacing the
dimensionality $d$ by $2\omega$, we calculate the integrals in the
domain of their convergence and then analytically continue the
result to $\omega \rightarrow d/2$. In spaces with even number of
dimensions, which we mainly consider in what follows, this gives
rise to the contribution $\SGamma_{{\rm div,\,log}}$ containing
the pole at $\omega =d/2$ and the term logarithmic in $m^{2}$
    \begin{equation}
    \SGamma_{{\rm div,log }}\!=\!\frac{1}{2(4\pi )^{d\over 2}}\int
    dx\sum\limits_{n=0}^{d/2}\!
    \frac{(-m^{2})^{{d\over 2}-n}}{\left({d\over2}
    -n\right) !}\left[\frac{1}{\omega -\frac{d}{2}}\!
    -\!\Gamma ^{\prime }\Big(
    \frac{d}{2}-n+1\Big)\!
    +\!\ln \frac{m^{2}}{4\pi \mu ^{2}}\right] a_{n}(x,x),       \label{dix}
    \end{equation}
where $\omega \rightarrow d/2.$ The pole corresponds to an
infinite ultraviolet renormalization of the terms proportional to
$a_{0},...,a_{d/2}$ in the original Lagrangian. Other terms in the
expansion (\ref{dewittact1}) are finite and give the infrared
contribution to the total action
    \begin{equation}
    \SGamma=\SGamma_{{\rm div,\ln }}-\frac{1}{2}
    \left( \frac{m^{2}}{4\pi}\right)^{d/2}
    \int dx\,\sum\limits_{n=d/2+1}^{\infty} \,
    \frac{\Gamma(n-d/2)}{(m^{2})^{n}}\,a_{n}(x,x).  \label{infdewitt}
    \end{equation}
The Schwinger-DeWitt coefficients $a_{n}(x,x)$ are the homogeneous
polynomials of dimensionality $2n$ in the units of inverse length, which are
built of $v(x)$ and its multiple derivatives . Therefore, on dimensional
grounds, they can be symbolically written down as
    \[
    a_{n}(x,x)\sim v^{k}(x)(\nabla ^{i}v^{j})(x),\,\,
    \]
where $i$ denotes an overall number of derivatives acting in all
possible ways on $j$ factors of $v(x)$, and $k$ powers of $v(x)$
stay undifferentiated. The positive integers $(k,j,i)$ are related
to $n$ as $2(k+j)+i=2n$. It is clear that the infinite series in (\ref%
{infdewitt}) represents the expansion in growing powers of the
following dimensionless quantities
    \begin{equation}
    \frac{v\left( x\right) }{m^{2}}\ll 1,\,\,\,\,
    \frac{\nabla ^{i}v(x)}{m^{2+i}}\ll 1,  \label{exppar}
    \end{equation}
which obviously should be much smaller than unity. Only in this
case the first few terms in the asymptotic series
(\ref{infdewitt}) are reliable.

Thus, the Schwinger-DeWitt expansion is applicable only in
theories with {\it small and slowly varying fields} as compared to
a big mass parameter. This expansion contains only local terms.
This is not surprising because all nonlocal effects, e.g.,
particle creation, are very small for heavy particles in a weak
external field and cannot be handled by this method. The
Schwinger-DeWitt technique can be easily extended to curved
spacetime and to theories with covariant derivatives built with
respect to an arbitrary fibre-bundle connection. In this case, the
perturbation potential $v(x)$ will also depend on the spacetime
curvature tensor and fibre-bundle curvatures (commutator of
covariant derivatives). The smallness of fields and their
derivatives includes the requirement of the smallness of these
curvatures and their derivatives as well. Despite its
universality, the Schwinger-DeWitt expansion becomes inefficient
when the ratios in (\ref{exppar}) become of the order of unity and
completely fails for massless fields. In the last case all
integrals over the proper time integral are infrared divergent.
This divergence has, of course, no physical meaning and is an
artifact of the approximation technique used.

There are two known ways to proceed with massless fields. One
possibility is the resummation of all terms which contain the
undifferentiated potential $V(x)$ in the local Schwinger-DeWitt
series (\ref{dewittexp}). They are summed up to form an exponent
similar to $e^{-sm^{2}}$
    \begin{equation}
    \Omega (s|\,x,x)=e^{-sV(x)}\sum\limits_{n=0}^{\infty }
    \tilde{a}_{n}(x,x)\,s^{n}.                 \label{moddewitt}
    \end{equation}
This method was suggested in \cite{MWZ}, where a regular technique
for the calculation of the modified Schwinger-DeWitt coefficients
$\tilde{a}_{n}(x,y)$ was also presented. The proper time integral
in (\ref{act2}) has now an infrared cutoff at $s\sim 1/V(x)$ and
in this case the effective action is similar to (\ref{dix}) -
(\ref{infdewitt}), where $m^{2}$ is replaced by $V(x)$ and
$a_{n}(x,x)$ by $\tilde{a}_{n}(x,x)$. It is convenient to write
this action as a sum of three terms
    \begin{equation}
    \SGamma=\SGamma_{{\rm div}}+\SGamma_{{\rm CW}}
    +\SGamma_{{\rm fin}},                     \label{totmodact}
    \end{equation}
where the divergent part is equal to
    \begin{equation}
    \SGamma_{{\rm div}}=\frac{1}{2(4\pi )^{\frac{d}{2}}}\int
    dx\sum\limits_{n=0}^{d/2}\frac{(-V)^{{d\over2}-n}}
    {\left({d\over2}-n\right)}\left[
    \frac{1}{\omega-{d\over 2}}-\Gamma^\prime
    \left({d\over 2}-n+1\right)-\ln 4\pi \right]
    \tilde{a}_{n}(x,x).                        \label{moddiv}
    \end{equation}
The pole part of this action coincides with that of (\ref{dix}) if
we take $m^{2}\to 0$ limit of (\ref{dix}). Actually, in this case,
only the term proportional to $a_{d/2}$ survives in $\SGamma_{{\rm
div,\,log}}$ and by virtue of the relation between twiddled and
untwiddled coefficients, namely,
    \begin{equation}
    a_{d/2}(x,x)=\sum\limits_{n=0}^{d/2}
    \frac{(-V)^{d/2-n}}{(d/2-n)!}\,
    \tilde{a}_{n}(x,x),\,\,\,  \label{rel}
    \end{equation}
the pole parts of (\ref{dix}) and (\ref{moddiv}) are the same. The
terms proportional to $\Gamma ^{\prime }(d/2-n+1)$ perform finite
renormalization of the local terms $V^{d/2-n}\,\tilde{a}_{n}.$ The
logarithmic terms of (\ref{dix}) are replaced in the modified
action (\ref{totmodact}) by
    \begin{equation}
    \SGamma_{{\rm CW}}=\frac{1}{2(4\pi )^{d/2}}
    \int dx\sum\limits_{n=0}^{d/2}%
    \frac{(-V)^{d/2-n}}{(d/2-n)!}\,
    \ln \frac{V}{\mu ^{2}}\,\,\tilde{a}_{n}=
    \frac{1}{2(4\pi )^{d/2}}\int dx\,
    \ln \frac{V}{\mu ^{2}}\,a_{d/2}.           \label{CW}
    \end{equation}
This is nothing but the space-time integral of the
Coleman-Weinberg effective potential. For instance, in four
dimensions the leading term is the original Coleman-Weinberg
effective potential, $V^{2}\ln (V/\mu^{2})/64\pi ^{2}$, while the
rest represents corrections due to the derivatives of $V(x)$.
Similarly to (\ref{infdewitt}), the finite part $\SGamma_{{\rm
fin}}$ is an infinite series
    \begin{equation}
    \SGamma_{{\rm fin}}=
    -\frac{1}{2}\int dx\,
    \left( \frac{V(x)}{4\pi }\right)^{d/2}
    \sum\limits_{n=d/2+1}^{\infty }\Gamma (n-d/2)\,
    \frac{\tilde{a}_{n}(x,x)}{V^{n}(x)}.  \label{modfin}
    \end{equation}
The modified Schwinger-DeWitt coefficients do not contain the
undifferentiated potential and the typical structure of the terms
entering $\tilde{a}_{n}(x,x)$ is $\nabla ^{m}V^{j}(x)$, where
$m+2j=2n$. Every $V$ here should be differentiated at least once
and therefore $m\geq j$. Thus the coefficients $\tilde{a}_{n}$ can
be symbolically written down as
    \begin{equation}
    \tilde{a}_{n}(x,x)\sim \sum\limits_{j=1}^{[2n/3]}
    \nabla ^{2n-2j}V^{j},                      \label{struc}
    \end{equation}
where the upper value of $j$ is the integer part of $2n/3$.

This perturbation theory is efficient as long as the potential is
slowly varying or bounded from below by a large positive constant,
so that
    \begin{equation}
    \frac{\nabla ^{2}V(x)}{V^{2}(x)}\ll 1,\,\,\,
    \frac{\left( \nabla V(x)\right) ^{2}}
    {V^{3}(x)}\ll 1,...\, .                   \label{parexp}
    \end{equation}
The case of bounded potentials reproduces the original
Schwinger-DeWitt expansion for nonvanishing mass. Therefore, let
us consider the potentials which vanish at spacetime infinity
$(|x|\rightarrow \infty )$. Namely, we assume the case of
power-like falloff
    \begin{equation}
    V(x)\sim \frac{1}{|x|^{p}},\,\,\,\,
    \nabla ^{m}V(x)\sim \frac{1}{|x|^{p+m}},
    \,\,\,\,|x|\rightarrow \infty          \label{potfall}
    \end{equation}
for some positive $p$. For such a potential terms of the
perturbation series (\ref{modfin}) behave as
    \begin{equation}
    \frac{\tilde{a}_{n}}{V^{n}}\sim
    \sum\limits_{j=1}^{[2n/3]}|x|^{(p-2)(n-j)}   \label{exppar1}
    \end{equation}
and thus decrease with $n$ only if $p<2$. For $p\geq 2,$ the
modified gradient expansion completely breaks down. It makes sense
only for slowly decreasing potentials of the form (\ref{potfall})
with $p<2$. In this case the potential $V(x)$ is not integrable
over the whole spacetime $\left( \int dx\,V(x)=\infty \right)$ and
moreover even the operation $(1/\Box )V(x)$ is not well
defined\footnote{For the convergence of the integral in
$\left(1/\Box \right) V$ the potential $V(x)$ should fall off at
least as $1/|x|^{3}$ in any spacetime dimension \cite{CPTII}.}.
Therefore the above restriction is too strong to account for many
interesting physical problems. In addition, similarly to
(\ref{infdewitt}), the asymptotic expansion (\ref{modfin}) is
completely local. It does not allow us to capture nonlocal
effects, which are exponentially small for potentials satisfying
(\ref{parexp}).

The way to overcome this difficulty and to take into account
nonlocal effects was suggested in the form of covariant
perturbation theory (CPT) \cite{CPTI,CPTII,CPTIII}. In this theory
the full potential $V(x)$ is treated as a perturbation and the
solution of the heat equation is found as a series in its powers.
From the viewpoint of the Schwinger-DeWitt expansion it
corresponds to an infinite resummation of all terms with a given
power of the potential and arbitrary number of derivatives. The
result reads as
    \begin{equation}
    {\rm Tr}\,K(s)\equiv \int dx\,K(s|\,x,x)=
    \sum\limits_{n=0}^{\infty }{\rm Tr}
    \,K_{n}(s),                         \label{CPT1}
    \end{equation}
where
    \begin{equation}
    {\rm Tr}\,K_{n}(s)=\int
    dx_{1}dx_{2}...dx_{n}\,
    F_{n}(s|\,x_{1},x_{2},...x_{n})%
    \,V(x_{1})V(x_{2})...V(x_{n}),  \label{Kn}
    \end{equation}
and the nonlocal form factors $F_{n}(s|\,x_{1},x_{2},...x_{n})$
were explicitly obtained in \cite{CPTI,CPTII,CPTIII}. It was shown
that at $s\rightarrow \infty $ the terms in this expansion behave
as
    \begin{equation}
    {\rm Tr}\,K_{n}(s)=
    O\left( \frac{1}{s^{d/2-1}}\right) ,\,\,\,
    n\geq 1,                                \label{CPTas}
    \end{equation}
and, therefore in spacetime dimension $d\geq 3$ the integrals in
(\ref{act2}) are infrared convergent
    \begin{equation}
    \int\limits^{\infty }\frac{ds}{s}\,
    O\left( \frac{1}{s^{d/2-1}}\right)
    <\infty .                 \label{infCPT}
    \end{equation}
In one and two dimensions this expansion for $\SGamma$ does not
exist except for the special case of the massless theory in curved
two-dimensional spacetime, when it reproduces the so-called
Polyakov action\cite{FrolVilk,Polyakov,CPTII}\footnote{This action
can be obtained by integrating the conformal anomaly \cite
{FrolVilk,Polyakov}.}. CPT should always be applicable whenever
$d\geq 3$ and the potential $V$ is sufficiently small\footnote{
The conditions of the smallness of the potential are exactly
opposite to those of (\ref{parexp}), e.g., $V^{2}/\nabla ^{2}V\ll
1$. However, this is true only as a rather rough estimate, because
CPT is a nonlocal perturbation theory and its actual smallness
``parameters'' are some nonlocal functionals of the potential.},
so that its effective action explicitly features analyticity in
the potential at $V=0$. Therefore, its serious disadvantage is
that this theory does not allow one to overstep the limits of
perturbation scheme and, in particular, discover non-analytic
structures in the action if they exist.

All this implies the necessity of a new approximation technique
that would allow us to overcome disadvantages of the above
methods. In the rest of this paper we develop such a technique,
involving further resummation of the perturbation series. We
develop an infrared improved perturbation theory for the heat
kernel and reveal new nonlocal and non-analytical structures in
the effective action.

\section{Resummation of proper time series}

We use the exponential ansatz for the function $\Omega (s|\,x,y)$
defined by Eq.(\ref{ansatz}):
    \[
    \Omega (s|\,x,y)=\exp \Big[-W(s|\,x,y)\Big].
    \]
Our goal is to develop an approximation technique for $W$ similar to CPT,
which is an alternative to the expansion in $s$. By virtue of (\ref{heateq})
and (\ref{incond}) the function $W(s|\,x,y)$ satisfies the equation
    \begin{equation}
    \frac{\partial W}{\partial s}+\frac{(x-y)^{\mu }}
    {s}\nabla _{\mu }W-\Box
    W=V-(\nabla W)^{2},                       \label{eqw}
    \end{equation}
with the initial condition
    \begin{equation}
    W(s=0|\,x,y)=0.  \label{iniW}
    \end{equation}
This equation is nonlinear in $W$ and we solve it by iteration, considering $%
(\nabla W)^{2}$ as a perturbation. For this purpose it is
convenient to rewrite (\ref{eqw}) - (\ref{iniW}) as an integral
equation. In Appendix A, it is shown that this integral equation
takes the following form
    \begin{equation}
    W(s|\,x,y)=s\int_{0}^{1}d\alpha \,e^{s\alpha (1-\alpha )
    \overset{-}{\Box }}\left( V(\bar{x})-
    (\nabla W(s\alpha |\,\bar{x},y))^{2}\right),      \label{intform}
    \end{equation}
where the operator $\overset{-}{\Box }$ acts on the argument
$\bar{x}\equiv \bar{x}(\alpha |x,y)=\alpha x+(1-\alpha )y.$ The
equation (\ref{intform}) can be easily solved if $(\nabla
W)^{2}\ll V$. As we will see later, this condition is satisfied
for a broad class of potentials $V$. The lowest order
approximation for $W$ is obtained by omitting $(\nabla W)^{2}$ in
eq. (\ref{intform})
    \begin{equation}
    W_{0}(s|\,x,y)=s\int_{0}^{1}d\alpha \,
    e^{s\alpha (1-\alpha )\overset{-}{\Box}}\left.
    V(\bar{x})\,\right| _{\bar{x}=\alpha x+(1-\alpha )y}.  \label{lowapp}
    \end{equation}
This is a linear but essentially nonlocal functional of the potential.
Further terms of the perturbation theory, $W_{n}=O\left[ (\nabla W_{0})^{n}%
\right] $, can be graphically represented by connected tree graphs with two
derivatives in the vertices, internal lines associated with the nonlocal
operator
    \begin{equation}
    f(-s\Box )=\int_{0}^{1}d\alpha \,
    e^{s\alpha (1-\alpha )\Box },  \label{oper}
    \end{equation}
and external lines given by (\ref{lowapp})\footnote{This graphical
interpretation should not be taken too literally because
integration over $\alpha $-parameter(s) involves also the argument
$\bar{x}\equiv \bar{x}(\alpha |x,y)$ of the potential.}. Note that
this connected graph structure arises in the exponential and when
expanded gives rise to the disconnected graphs. In context of the
heat kernel expansion this property was observed in
\cite{connected}. Resummation of the perturbation series in $V$
explicitly features exponentiation of the quantities containing
only connected graphs. Here we showed how this exponentiated set
of connected graphs directly arises from the solution of the
simple nonlinear equation (\ref{eqw}). The ''propagator''
(\ref{oper}) was worked out within the covariant perturbation
theory in \cite {CPTI,CPTII,CPTIII} and was also obtained in
\cite{MWZ} by direct summation of gradient series.

At this stage the efficiency of the connected graph expansion is
not yet obvious. Crudely, it runs in powers of the dimensionless
quantity $\left( sf(-s\Box )\nabla \right) ^{2}V(x)$ which, at
least naively, should be small for slowly varying or/and small
potentials. Apart from this, infrared properties of the effective
action strongly depend on the lowest order approximation for $W$
(\ref{lowapp}). The effective action involves only diagonal
elements of the two-point function $W(s|\,x,y),$ which look much
simpler than (\ref{lowapp})
    \begin{equation}
    W_{0}(s|\,x,x)=s\int_{0}^{1}d\alpha \,
    e^{s\alpha (1-\alpha )\Box }V(x).       \label{coinarg}
    \end{equation}
Note that, at small $s,$ the function $W_{0}$ can be expanded as $%
W_{0}=sV+O\left( s^{2}\right) .$ The only term with
undifferentiated potential entering $W_{0}$ is linear in $s,$
while all other terms contain derivatives of the potential $V.$
The same is also true for the exact $W,$ which differs from
$W_{0}$ by higher powers of the differentiated potential. This
completely agrees with the modified gradient expansion discussed
in Sect.2. However, the expression (\ref{coinarg}) directly
involves the nonlocal operator and its late time behavior is very
different from that naively expected in the modified gradient
expansion and in CPT.

To show that, let us first find the coordinate representation of
the operator (\ref{oper}), that is, the kernel
$f(-s\Box)\delta^{(d)}(x,y)$. Using a well-known result for the
exponentiated $\Box $-operator
    \begin{equation}
    e^{s\alpha (1-\alpha )\Box }\delta^{(d)}(x,y)=\frac{1}{(4\pi s\alpha
    (1-\alpha ))^{d/2}}\exp \left( -\frac{|x-y|^{2}}{4s\alpha (1-\alpha )}%
    \right),  \label{expbox}
    \end{equation}
one can write
    \begin{eqnarray}
    &&\int_{0}^{1}d\alpha \,
    e^{s\alpha(1-\alpha)\Box}
    \delta ^{(d)}(x,y) \nonumber \\
    &&\qquad\qquad\quad
    =\frac{e^{-|x-y|^{2}/2s}}{(4\pi s)^{d/2}}
    \int_{0}^{\infty}d\beta \,
    \frac{(1+\beta )^{d-2}}{\beta ^{d/2}}
    \exp \left[ -\frac{|x-y|^{2}}{4s}
    \left( \frac{1}{\beta }+\beta \right) \right] ,  \label{kern1}
    \end{eqnarray}
where we have changed the integration variable,
$\alpha=\beta/(1+\beta )$, $0\leq \beta <\infty $. For $d\geq 3,$
the integral can be easily calculated and the result is expressed
as a sum of McDonald functions of the argument  $|x-y|^{2}/2s$
    \begin{equation}
    f(-s\Box )\delta ^{(d)}(x,y)=
    \frac{2e^{-|x-y|^{2}/2s}}{(4\pi s)^{d/2}}%
    \sum\limits_{k=1}^{d-1}C_{k-1}^{d-2}
    K_{k-\frac{d}{2}}\left(
    |x-y|^{2}/2s\right) ,  \label{kern2}
    \end{equation}
where $C_{k-1}^{d-2}$ are the binomial coefficients. For very
large $s$, the argument of $K_{k-\frac{d}{2}}(z)$ is small. Using
the asymptotics: $K_{\nu }(z)\simeq \Gamma (|\nu |)(2/z)^{|\nu
|}/2,\,\,\,z\rightarrow 0$, we find that at large $s$ the form
factor is dominated by the following term
    \begin{equation}
    f(-s\Box )\delta ^{(d)}(x,y)
    =\frac{1}{2s}\frac{\Gamma (d/2-1)}{\pi^{d/2}|x-y|^{d-2}}
    +O\left( \frac{1}{s^{2}}\right) .  \label{kern3}
    \end{equation}
Taking into account the fact that
    \begin{equation}
    \frac{1}{\Box }\delta ^{(d)}(x,y)=
    -\frac{\Gamma (d/2-1)}{4\pi^{d/2}|x-y|^{d-2}},  \label{invbox}
    \end{equation}
we finally obtain:
    \begin{equation}
    W_{0}(s|\,x,x)=sf(-s\Box )V=
    -\frac{2}{\Box }V(x)+O\left( \frac{1}{s}\right) . \label{asW0}
    \end{equation}
This behavior agrees with the formal asymptotics found by the
Laplace method in \cite{CPTII}. Therefore, at
$s\rightarrow\infty$, the function $W_{0}$ approaches a constant.
The nonlocal functional (\ref{asW0}) is well-defined for $d\geq 3$
only if the potential $V$ vanishes fast enough at $|x|\rightarrow
\infty$.

For splitted arguments the asymptotics of $W_{0}(s|\,x,y)$ is more
intricate. In this case the form factor (\ref{oper}) no longer
arises as a whole because the integration parameter $\alpha $
appears in eq.(\ref{lowapp}) also in the argument $\bar{x}=\alpha
x+(1-\alpha )y$ of the potential $V(\bar{x})$. Applying the
Laplace method, one can show that the integral (\ref{lowapp}) is
dominated by the contribution of the end points, $\alpha =0$ and
$\alpha =1$. These contributions are different because
$\bar{x}(\alpha =0)=y$ and $\bar{x}(\alpha =1)=x$, whence
    \begin{equation}
    W_{0}(s|\,x,y)=-\frac{1}{\Box }V(x)
    -\frac{1}{\Box }V(y)
    +O\left( \frac{1}{s}\right) .  \label{assplitarg}
    \end{equation}
Substituting this expression in (\ref{intform}) and solving the
integral equation by iterations one can find the late time
asymptotics $W_\infty(x,y)$ for the exact $W(s|\,x,y)$
    \begin{equation}
    W(s|\,x,y)=W_\infty(x,y)
    +O\left( \frac{1}{s}\right)        \label{exas1}
    \end{equation}
as a nonlocal gradient series
    \begin{equation}
    W_\infty(x,y)=
    -\frac{1}{\Box }V(x)+
    \frac{1}{\Box }\left( \frac{1}{\Box }
    \nabla V(x)\right)^{2}+...
    +(x\leftrightarrow y).                \label{exas}
    \end{equation}
It is remarkable, however, that this series can be ''summed up''
and the nonlocal expression for $W_{\infty }(x,y)$ can be found
exactly in terms of the Green's function of the original operator
$\widehat{F}=-\Box +V$.

\section{Late time asymptotics of the heat kernel}

By substituting the late time ansatz (\ref{exas1}) in equation
(\ref{eqw}), it is easy to see that the first two terms in its
left hand side vanish at $s\rightarrow \infty $, while the rest
reduce to the equation for $W_{\infty }(x,y)$
    \begin{equation}
    (\Box -V)\,e^{-W_{\infty }(x,y)}=0.  \label{eqWinf}
    \end{equation}
Despite the positivity of the operator $-\Box +V$ with Dirichlet
boundary conditions at $|x|\to\infty$, this equation admits
non-trivial solutions. In fact, $e^{-W_{\infty }(x,y)}$ does not
have to go to zero at infinity\footnote{The boundary condition
$K(s|\,x,y)\rightarrow 0$ at $|x|\rightarrow \infty$ is enforced
by the gaussian factor in (\ref{ansatz}), even for nonvanishing
finite $\Omega(s|\,x,y)=\exp(-W(s|\,x,y)$.}. In view of the
iterative solution (\ref{exas}) it should tend to some unknown
function of $y$
    \begin{equation}
    e^{-W_{\infty }(x,y)}\rightarrow C(y),\,\,\,
    |x|\rightarrow \infty .            \label{bound1}
    \end{equation}
The equation (\ref{eqWinf}) with this boundary condition is then
solved by
    \begin{equation}
    e^{-W_{\infty }(x,y)}=C(y)\,\Phi (x),  \label{ans}
    \end{equation}
if the new function $\Phi (x)$ satisfies the equation
    \begin{equation}
    (\Box -V)\,\Phi (x)=0,  \label{eqF}
    \end{equation}
and $\Phi (x)\rightarrow 1$ at $|x|\rightarrow\infty$. The
solution of this problem for $\Phi(x)$ is uniquely determined in
terms of the Green's function (\ref{GR})
    \begin{equation}
    \Phi (x)=1+\frac{1}{\Box-V}\,V(x).  \label{solF}
    \end{equation}
This function of $x$ is a nonlocal and essentially nonlinear
functional of the potential, and it will play a very important
role in what follows.

The heat kernel is symmetric in its arguments $x,y$ and,
therefore, the unknown function $C(y)$ should coincide with
$\Phi(y)$. Thus, finally we obtain the following exact late time
asymptotics
    \begin{equation}
    e^{-W_{\infty }(x,y)}=\Phi (x)\,\Phi (y).  \label{finalas}
    \end{equation}
Expanding $W_{\infty }(x,y)$ in powers of the potential $V$ one
gets
    \begin{eqnarray}
    W_{\infty }(x,y) &=&-\ln \Phi (x)-\ln \Phi (y)   \nonumber\\
    &=&-\frac{1}{\Box }V(x)
    -\frac{1}{\Box }V\frac{1}{\Box }V(x)
    +\frac{1}{2}\left( \frac{1}{\Box }V(x)\right)^{2}+...
    +(x\leftrightarrow y).                              \label{exexp}
    \end{eqnarray}
The first term here is in agreement with the perturbative
asymptotics (\ref{exas}). However, beyond that, the iterative
solution (\ref{exas}) seems to be in contradiction with
(\ref{exexp}). The series (\ref{exas}) runs in powers of the
differentiated potential, while the expansion (\ref{exexp})
contains only undifferentiated potential. However, integration by
parts in the second term of (\ref{exas}) exactly reproduces the
second and third terms of (\ref{exexp}). Thus, both expansions are
equivalent, but the first one reveals more explicitly the gradient
of the potential as a small parameter, while in (\ref{exexp}) the
smallness is a result of non-trivial cancellations between
different terms.

Finally we write down the exact late time asymptotics for the heat kernel,
advocated in Introduction,
    \begin{equation}
    K(s|\,x,y)=\frac{1}{(4\pi s)^{d/2}}\,
    \Phi (x)\Phi (y),\,\,\,s\rightarrow\infty .  \label{hkas}
    \end{equation}
Its heuristic interpretation is rather transparent. The heat kernel can be
decomposed in the series
    \begin{equation}
    K(s|\,x,y)=\sum\limits_{\lambda }
    e^{-\lambda s}\Phi _{\lambda }(x)\,
    \Phi_{\lambda }(y),  \label{eigenexp}
    \end{equation}
where $\lambda$ and $\Phi _\lambda$ are, respectively, the
eigenvalues and eigenfunctions of the operator $\widehat{F}=-\Box
+V(x)$. Since $\widehat{F}$ is a positive-semidefinite operator,
only the lowest eigenmode with $\lambda =0$ survives in this
expression in the limit $s\rightarrow \infty$. The appropriate
eigenfunction satisfies the equation $\widehat{F}\Phi _{\lambda
=0}=0$, which coincides with (\ref{eqF}) and therefore $\Phi
_{\lambda =0}(x)=\Phi (x)$. The spectrum of the operator is
continuous and the eigenmodes are not square integrable
$\left(\Phi _{0}(x)\rightarrow 1\text{ at } |x|\rightarrow \infty
\right)$. This is why the integral over the spectrum, denoted
above by $\sum_{\lambda }$, yields within the steepest decent
approximation the power-like asymptotics, $1/s^{d/2}$, rather than
the exponential one. Of course, these arguments are not very
rigorous. The zero mode $\Phi (x)$ with unit boundary condition at
infinity, does not even belong to the continuous spectrum of modes
normalized to the delta function. Nevertheless, as we have seen,
this particular mode gives the leading contribution to the late
time asymptotics of the heat kernel.

If we want to calculate the contribution to the effective action
due to the late time behavior of the heat kernel we need its
functional trace -- spacetime integral of the coincidence limit
$K(s|\,x,x)$. Unfortunately, the expression (\ref{hkas}) cannot be
used directly to calculate ${\rm Tr}\,K(s)$ for a given $s$. Point
is that this asymptotic expression taken at a fixed large $s$ is
applicable only for $|x|^{2}<s$ and fails at $|x|^{2}\gg
s$\footnote{This follows from the derivation of (\ref{hkas})
above, which is based on discarding the second term of
eq.(\ref{eqw}) linearly growing in $(x-y)/s$.}. At the same time,
when calculating the trace we have to integrate over the whole
spacetime up to $|x|\rightarrow \infty $ and therefore need the
heat kernel behavior for $|x|^2\gg s$. The attempt to disregard
this subtlety and integrate the coincidence limit of (\ref{hkas})
over $x$ results in a poorly defined quantity -- the spacetime
integral strongly divergent at infinity. Nevertheless, one can use
the expression (\ref{hkas}) to find ${\rm Tr}\,K(s)$ with the aid
of the following somewhat subtler procedure.

First let us write the variational relation
    \begin{equation}
    \delta \,{\rm Tr}K(s)=-s{\rm Tr}\,
    \big(\delta V\,K(s)\big)
    =-s\int dx\,\delta V(x)\,K(s|x,x),  \label{var}
    \end{equation}
where, of course, $K(s)=\exp [s(\Box -V)]$. Then it follows that
    \begin{equation}
    \frac{\delta \,{\rm Tr}K(s)}{\delta V(x)}
    =-sK(s|x,x).  \label{vareq}
    \end{equation}
Substituting $K(s|x,x)$ from (\ref{hkas}) in the right hand side of this
relation we obtain the following functional differential equation
    \begin{equation}
    \frac{\delta \,{\rm Tr}K(s)}{\delta V(x)}=
    -\frac{s}{(4\pi s)^{d/2}}\,\Phi^{2}(x).  \label{feq}
    \end{equation}
This equation satisfies the intergrability condition, because the variational
derivative
    \begin{equation}
    \frac{\delta\Phi^2(x)}{\delta V(y)}=2\Phi(x)\Phi(y)
    \frac1{\Box-V}\delta(x,y)
    \end{equation}
is symmetric in $x$ and $y$. Therefore (\ref{feq}) can be
integrated to determine ${\rm Tr}K(s)$. The solution subject to
the obvious boundary conditions at $V=0$ reads
    \begin{equation}
    {\rm Tr}\,K(s)=\frac{1}{(4\pi s)^{d/2}}
    \int dx\,\Big( 1-sV\Phi \Big),
    \,\,\, s\to\infty.                     \label{sol}
    \end{equation}
One can easily check that this expression satisfies the equation (\ref{feq}).

It is remarkable that in the covariant perturbation theory the
leading and next subleading (in $s$) terms of the heat kernel
trace can be explicitly calculated for $s\rightarrow\infty $ in
every order of expansion in powers of $V$. The corresponding
infinite series can be explicitly summed up to yield essentially
nonlocal and nonlinear in $V(x)$ expression for ${\rm Tr}\,K(s)$.
This is done in Appendix B to the first subleading order
inclusive. The following very simple and concise result reads as
    \begin{eqnarray}
    {\rm Tr}\,K(s)=\frac1{(4\pi s)^{d/2}}
    \int dx\,\left\{1-sV\Phi
    -2\nabla_\mu\Phi\frac1{\Box-V}\nabla^\mu\Phi
    +O\left(\frac1s\right)\right\}        \label{LATETR}
    \end{eqnarray}
in terms of the function $\Phi(x)$ and its derivatives. As we see,
it exactly reproduces the leading order term of (\ref{sol}),
$O(s/s^{d/2})$, and also gives a nontrivial
$O(1/s^{d/2})$-correction. Below we use this asymptotics for
obtaining new types of the nonlocal effective action.

\section{Effective action}

The functional trace of the heat kernel is everything we need for
the calculation of the effective action. Unfortunately, only its
asymptotics are known.  Namely, at small $s$ one can use the
modified gradient expansion (\ref{ansatz}), (\ref{moddewitt}) and
at large $s$ -- the nonlocal and nonlinear expression
(\ref{LATETR}). The goal of this section is to unify both of these
approximations to get the expression for the effective action
which would incorporate both the ultraviolet and infrared
properties of the theory. The calculation will be explicitly done
in four dimensional case. The generalization to other dimensions
$d>2$ is straightforward.

The key idea is to replace ${\rm Tr}\,K(s)$ in (\ref{act2}) by some
approximate function ${\rm Tr}\,\bar{K}(s)$ such that the integral over $s$
    \begin{equation}
    \bar{\SGamma}=-\frac{1}{2}\,\int
    \frac{ds}{s}{\rm Tr}\,\bar{K}(s)             \label{act1a}
    \end{equation}
becomes explicitly calculable. The difference ${\rm
Tr}\,K(s)-\bar{K}(s)$ can then be treated as a perturbation.
Certainly, the efficiency of this procedure very much depends on
the successful choice of $\bar{K}(s)$. Here we exploit the
simplest possibility -- namely, let us take two simple functions
${\rm Tr}\,\bar{K}_{<}(s)$ and ${\rm Tr}\,\bar{K}_{>}(s)$, which
coincide with the leading asymptotics of ${\rm Tr}\,K(s)$ at
$s\rightarrow 0$ and $s\rightarrow \infty$ and use them to
approximate ${\rm Tr}\,K(s)$ respectively at $0\leq s\leq
s_{\ast}$ and $s_{\ast }\leq s<\infty$. In turn, the value of
$s_{\ast}$ will be determined from the requirement that these two
functions match at $s_{\ast}$. This will guarantee the
stationarity of $\bar{\SGamma}$ with respect to the choice of
$s_{\ast}$, $\partial \bar{\SGamma}/\partial s_{\ast}=0$. We will
discuss the justification of this procedure a little later, while
now let us proceed with the calculation of $\bar{\SGamma}$.

At small $s$ we use the lowest order term of the modified gradient
expansion
    \begin{equation}
    {\rm Tr}\,K_{<}(s)=\frac{1}{(4\pi s)^{2}}
    \int dx\,\exp \left( -Vs\right),\,\,\,s<s_{\ast }  \label{ass1}
    \end{equation}
and disregard all terms containing derivatives of the potential
$V$. The rest of the asymptotic series (\ref{moddewitt}),
containing $\tilde{a}_{n}$, will be treated by perturbations.
Correspondingly, at large $s>s_{\ast }$ we use the late time
asymptotics (\ref{sol})
    \begin{equation}
    {\rm Tr}\,K_{>}(s)=\frac{1}{(4\pi s)^{2}}
    \int dx\,(1-sV\Phi),\,\,\,s>s_{\ast }.  \label{lass}
    \end{equation}

The requirement of stationarity of $\bar{\SGamma}$ with respect to
$s_{\ast}$ leads to the equation
    \begin{equation}
    \int dx\,\exp \left( -Vs_{\ast }\right)
    =\int dx\,(1-s_{\ast }V\Phi ),          \label{eqfors}
    \end{equation}
which determines the value of $s_{\ast}$ as some nontrivial
functional of the potential, $s_{\ast }=s_{\ast }[\,V(x)\,]$.
Unfortunately this functional is not calculable explicitly in
general, but nevertheless, as we will see below, it can be
obtained for two rather broad classes of potentials. The action
(\ref{act1a}) can be written down as a sum of two contributions
    \begin{equation}
    \bar{\SGamma}=\SGamma_{<}+\SGamma_{>}=
    -\frac{1}{2}\int\limits_{0}^{\infty }
    \frac{ds}{s}{\rm Tr}\,K_{<}(s)
    -\frac{1}{2}\int\limits_{s_{\ast }}^{\infty }
    \frac{ds}{s}\left( {\rm Tr}\,K_{>}(s)
    -{\rm Tr}\,K_{<}(s)\right) .                   \label{ll}
    \end{equation}
The first integral here has already been calculated and is given
by the sum of the expressions (\ref{moddiv}) and (\ref{CW}) with
$\tilde{a}_{0}=1$ and $\tilde{a}_{n}=0,\,n\geq 1$, which in our
particular case give rise to
    \begin{eqnarray}
    &&\SGamma_{<}=\SGamma_{{\rm div}}
    +\SGamma_{{\rm CW}}  \nonumber \\
    &&\qquad \equiv \frac{1}{64\pi ^{2}}
    \int dx\,\left[ \left( -\frac{1}{2-\omega}
    +2{\bf C}-3-\ln 4\pi \right) V^{2}
    +V^{2}\ln \frac{V}{\mu ^{2}}\,\right],
    \,\,\,\omega \rightarrow 2,                   \label{uv}
    \end{eqnarray}
where ${\bf C}=0,577...$ is the Euler's constant. The first term
here is responsible for the renormalization of the original action
and the second one is just the Coleman-Weinberg potential. The
second integral in (\ref{ll}) can also be calculated exactly.
Integrating by parts and taking into account (\ref{eqfors}), we
obtain
    \begin{eqnarray}
    &&\SGamma_{>}\equiv
    -\frac{1}{2}\int\limits_{s_{\ast }}^{\infty }
    \frac{ds}{s}\,\left( {\rm Tr}\,K_{>}(s)
    -{\rm Tr}\,K_{<}(s)\right)       \nonumber \\
    &&\qquad \qquad \qquad \qquad
    =\frac{1}{64\pi ^{2}}\int dx\,
    \left[ \,\frac{V\Phi }{s_{\ast }}
    -\frac{Ve^{-s_{\ast }V}}{s_{\ast }}
    +V^{2}\Gamma (0,s_{\ast}V)\,\right],  \label{ir}
    \end{eqnarray}
where $\Gamma (0,x)$ is an incomplete gamma function, $\Gamma
(0,x)=\int_{x}^{\infty }\,dt\,t^{-1}e^{-t}$, with the following asymptotics
    \begin{equation}
    \Gamma (0,x)\sim \left\{
    \begin{array}{lc}
    \ln \displaystyle{\frac{1}{x}}-{\bf C}, &\,\,\,\,x\ll 1, \\
    \displaystyle{\frac1x}\,e^{-x}, &\,\,\,\, x\gg 1.
    \end{array}\right.                   \label{gass}
    \end{equation}

Further steps strongly depend on the class of potentials, for
which the consistency of the piecewise approximation (\ref{ass1})
- (\ref{lass}) should be carefully analyzed.

\subsection{Small potential}

The approximation (\ref{ass1}) - (\ref{lass}) is efficient only if
the ranges of validity of two asymptotic expansions (respectively
for small and big $s$) overlap with each other and the point
$s_{\ast}$ belongs to this overlap. In this case the corrections
due to the deviation of ${\rm Tr}\,\bar{K}(s)$ from the exact
${\rm Tr}\,K(s)$ are uniformly bounded everywhere and one can
expect that (\ref{act1a}) would give a good zeroth-order
approximation to an exact result. Below we show that this
necessary requirement can be satisfied at least for two rather
wide classes of potentials $V(x)$.

The modified gradient expansion is well applicable in the overlap
range of the parameter $s$ if
    \begin{equation}
    s\nabla \nabla V\ll V,  \label{5.1}
    \end{equation}
(cf. Eq.(\ref{parexp}) with $s$ replaced by effective cutoff
$s=1/V$) and the applicability of the large $s$ expansion in the
same domain reads as
    \begin{equation}
    s\int dx\,V\Phi \gg \int dx\,
    \nabla _{\mu }\Phi \frac{1}{V-\Box }
    \nabla^\mu\Phi,                                  \label{5.4}
    \end{equation}
which means that the subleading term (quadratic in
$\nabla_\mu\Phi$) of the late time expansion (\ref{LATETR}) is
much smaller than the leading second term.

To implement these requirements, let us make some simplifying assumptions.
Instead of the power law falloff, assume that $V(x)$ has a compact support
of finite size $R$
    \begin{equation}
    V(x)=0,\,\,\,|x|\geq R.
    \end{equation}
Let us also assume that the characteristic magnitude of the potential inside
its support is given by $V_{0}$. Then the estimate for the derivatives is
obvious
    \begin{equation}
    \nabla \nabla V\sim \frac{V_{0}}{R^{2}}
    \end{equation}
and (\ref{5.1}) reads as $sV_{0}/R^{2}\ll V_{0}$, that is
    \begin{equation}
    s\ll R^{2}.                      \label{a2}
    \end{equation}
To find out what does the criterion (\ref{5.4}) mean let us make a
simplifying assumption, namely, that the potential $V$ is small.
In this case it can be disregarded in the Green's functions and
$1/(\Box-V)$ can be replaced by $1/\Box $. Therefore the following
estimates hold
    \begin{eqnarray}
    &&\frac{1}{\Box -V}V(x)\sim \int_{|y|\leq R}dy\,
    \frac{1}{|x-y|^{d-2}}%
    V(y)\sim \frac{1}{R^{d-2}}R^{d}V_{0}
    \sim V_{0}R^{2},                  \nonumber\\
    &&\int dx\,V\Phi \sim V_{0}R^{d}, \nonumber\\
    &&\int dx\,\nabla_\mu\Phi
    \frac{1}{V-\Box }\nabla ^{\mu }\Phi
    \simeq V_{0}^{2}R^{d+4}.                             \label{est}
    \end{eqnarray}
Roughly, every Green's function gives the factor $R^{2}$, every
derivative -- $1/R$, integration gives the volume of compact
support $R^{d}$, etc. Applying these estimates to eq. (\ref{5.4})
we get $sV_{0}R^{d}\gg V_{0}^{2}R^{d+4}$, whence
    \begin{equation}
    s\gg V_{0}R^{4}.
    \end{equation}
Combining this with (\ref{a2}) one gets the following range of
overlap of our asymptotic expansions
    \begin{equation}
    R^{2}\gg s\gg V_{0}R^{4}.  \label{a5}
    \end{equation}
It immediately follows from here that this overlap domain is not empty only
if
    \begin{equation}
    V_{0}R^{2}\ll 1.  \label{sp}
    \end{equation}
Moreover, the assumption of disregarding the potential in the
Green's function is also justified in this case since $V\sim
V_{0}\ll 1/R^{2}\sim \Box $. In other words this bound means that
the potential is small in units of the inverse size of its compact
support.

Now let us check whether $s_{\ast }$ introduced above belongs to
the overlap domain (\ref{a5}). Note that if it is really so, then
$s_{\ast}V$ in Eq.(\ref{eqfors}) is much smaller than unity
because in the overlap range one has $sV\sim sV_0\ll R^2 V_0\ll
1$. Hence the exponent in the left hand side of (\ref{eqfors}) can
be expanded in powers of $s_{\ast }V$, and the resulting equation
for $s_{\ast}$ becomes\footnote{Note that the quadratic term
should be retained in the expansion of $e^{-s_*V}$ if we want to
get a nontrivial solution for $s_{\ast}$.}
    \begin{equation}
    \int dx\,\left( 1-s_{\ast }V
    +\frac{s_{\ast }^{2}}{2}V^{2}
    +O\left( \left(s_{\ast}V\right)^{3}\right) \right)
    =\int dx\,\left( 1-s_{\ast }V\Phi\right)    \end{equation}%
Its solutions has the following form:
    \begin{equation}
    s_{\ast }\simeq 2\,\frac{\int dx\,
    V(1-\Phi )}{\int dx\,V^{2}}=2\,\frac{\int dx\,
    V\frac{1}{V-\Box }V}{\int dx\,V^{2}}.
    \end{equation}
Taking into account the estimates (\ref{est}) we see that the
point $s_{\ast}\sim R^{2}$ belongs to the upper edge of the
interval (\ref{a5}). Late time expansions is fairly well satisfied
here, but the small $s$ expansion is on the verge of breakdown. At
this level of generality it is hard to overstep the uncertainty of
this estimate. There is a hope that numerical coefficients in more
precise estimates (with concrete potentials) can be large enough
to shift $s_{\ast}$ to the interior of the interval (\ref{a5})
and, thus, make our approximation completely reliable.

Bearing in mind all these reservations let us proceed with the calculation
of the effective action. Using the small $x$ asymptotics (\ref{gass}) in the
expression (\ref{ir}) we get
\begin{equation}
\SGamma_{>}\simeq \frac{1}{64\pi ^{2}}\int d^4x\,\left[ \,-V\frac{1-\Phi }{%
s_{\ast }}+V^{2}\left( \ln \frac{1}{s_{\ast }V}-{\bf C}+1\right) \,\right] .
\end{equation}%
It is interesting to note that in the whole action $\bar{\SGamma}=\SGamma%
_{<}+\SGamma_{>}$ the Coleman-Weinberg term disappears and the final answer
reads
\begin{eqnarray}
&&\bar{\SGamma}\simeq \frac{1}{64\pi ^{2}}\left( -\frac{1}{2-\omega }+{\bf C}%
-2-\ln 4\pi \right) \int d^4x\,V^{2}  \nonumber \\
&&\quad +\frac{1}{64\pi ^{2}}\int d^4x\,V^{2}\ln \Big(\int dx\,V^{2}\Big)-%
\frac{1}{64\pi ^{2}}\int d^4x\,V^{2}\ln \Big(\int dx\,V\frac{\mu ^{2}}{V-\Box }%
V\Big).  \label{EAsmallV}
\end{eqnarray}%
The first term here differs from $\SGamma_{{\rm div}}$ in (\ref{uv}) by a
finite renormalization of the local $V^{2}$-term, while the two other terms
have entirely new nonlinear and nonlocal structure advocated in the
Introduction. The ultraviolet renormalization mass parameter $\mu ^{2}$
makes the argument of the second logarithm dimensionless -- it plays the
same role as for the Coleman-Weinberg potential, but now it enters the new
essentially nonlocal structure.

It is natural that the original Coleman-Weinberg term for the case
of small potentials (\ref{sp}) gets replaced by the other
qualitatively new nonlocal structure. Potentials which are small
in units of the inverse size of their support are qualitatively
very different from nearly constant potentials for which the
Coleman-Weiberg action was originally derived. In the case of
small potentials spacetime gradients dominate over their magnitude
and, therefore, one should not expect that the Colman-Weiberg term
would survive the inclusion of nonlocal structures.

\subsection{Big potential}

Remarkably, the case of the small potentials (\ref{sp}) is not the
only one when one can find a non-empty domain of overlap where
both asymptotics for ${\rm Tr}\,K(s)$ are applicable. Namely, the
opposite case of big potentials (in units of the inverse size of
their support)
    \begin{equation}
    V_{0}R^{2}\gg 1,  \label{bp}
    \end{equation}
is equally good. The key observation here is that in this case the
kernel of the Green's function $1/(\Box -V)$ can be replaced
within the compact support of $V$ by $-1/V$ ($\Box \sim 1/R^{2}\ll
V_{0}\sim V$) and correspondingly
    \begin{eqnarray}
    &&\frac{1}{\Box -V}V(x)
    \sim -\frac{1}{V}V=-1,  \label{5.12a} \\
    &&\int dx\,\nabla _{\mu }
    \Phi \frac{1}{V-\Box }
    \nabla ^{\mu }\Phi \simeq
    \frac{R^{4}}{V_{0}R^{2}}.  \label{5.12}
    \end{eqnarray}
Therefore, the criterion of applicability of the late time
expansion (\ref {5.4}) becomes $s\gg 1/V_{0}^{2}R^{2}$. Together
with (\ref{a2}) it yields the new overlap range
    \begin{equation}
    R^{2}\gg s\gg \frac{1}{V_{0}^{2}R^{2}}  \label{bpint}
    \end{equation}
which is obviously not empty if the potential satisfies (\ref{bp}).

To find $s_{\ast}$ in this case we have to solve the equation
(\ref{eqfors}) for the case when $s_{\ast }V$ is not anymore a
small quantity. Since $V$ is big, the exponent in (\ref{eqfors})
can be replaced by zero inside the compact support,
$\exp(-s_{\ast}V(x))\sim 0$, $|x|\leq R$, and by one outside of it
where the potential vanishes, $\exp (-s_{\ast }V(x))\sim 1$,
$|x|>R$. Rewriting the integrals in both sides of the equation
(\ref{eqfors}) as a sum of contributions of $|x|\leq R$ and
$|x|>R$, we see that the contribution of the non-compact domain
gets cancelled and the equation becomes
    \begin{equation}
    s_{\ast }\int\limits_{|x|\leq R}dx\,
    V\Phi \simeq \int\limits_{|x|\leq R}dx.
    \end{equation}
Then it follows that $s_{\ast}$ is approximately given by the
inverse of the function $V\Phi (x)$ {\em averaged} over the
compact support of the potential
    \begin{eqnarray}
    &&s_{\ast }\simeq \frac{1}{\big<V\Phi \big>}, \\
    &&\big<V\Phi \big>\equiv
    \frac{\int\limits_{|x|\leq R}\!dx\,V\Phi }
    {\int\limits_{|x|\leq R}\!dx}.
    \end{eqnarray}
A qualitative estimate of $\big<V\Phi \big>\sim V_{0}$ implies
that $s_{\ast}\sim 1/V_{0}$ and it belongs to the middle of the
interval (\ref{bpint}). This makes the case of big potentials
fairly consistent. On the other hand, the value of $\Phi (x)$ is
close to zero inside the potential support (see (\ref{5.12a})), so
most likely the estimate for $\big<V\Phi \big>$ is smaller by at
least one power of the quantity $1/V_{0}R^{2}$, which is the basic
dimensionless small parameter in this case. Therefore the
magnitude of $s_{\ast}$ becomes bigger by one power of
$V_{0}R^{2}$, $s_{\ast }\simeq R^{2}$, which is again near the
upper boundary of the overlap interval (\ref{bpint}). Similarly to
the small potential case, a more rigorous analyses is needed
(maybe for more concretely specified potentials) to account for
subtle edge effects at the boundary of compact support, which
might shift the value of $s_{\ast }$ to a safe region inside
(\ref{bpint}).

With the above estimate for $s_{\ast }\sim R^{2}$ the magnitude of
$s_{\ast}V$ in the expression for the infrared part of the
effective action (\ref{ir}) becomes big, $s_{\ast }V\sim
s_{\ast}V_{0}\sim V_{0}R^{2}\gg 1$, and we use the big $x$
asymptotics in (\ref{gass}) to get the contribution
    \begin{equation}
    \SGamma_{>}\simeq \frac{1}{64\pi ^{2}s_{\ast }}
    \int d^4x\,V\Phi =\frac{1}{64\pi ^{2}}\int\limits_{|x|\leq R}
    \!dx\,\big<V\Phi \big>^{2}.
    \end{equation}
In this case the Coleman-Weinberg term is not cancelled in
complete agreement with what we would expect for big potentials
and the final result reads
    \begin{equation}
    \bar{\SGamma}=\SGamma_{{\rm div}}
    +\SGamma_{{\rm CW}}+\frac{1}{64\pi ^{2}}
    \int\limits_{|x|\leq R}\!d^4x\,\big<V\Phi \big>^{2}.
    \end{equation}

\section{Comments}

We developed a new technique for the calculation of late time
asymptotics of the heat kernel and its functional trace. Using
these asymptotics we found previously unknown essentially nonlocal
and nonperturbative contributions to the effective action for two
large classes of potentials with compact supports. Therefore, the
generalization of these results to potentials with power-law
falloff, which would imply subtler analyses, deserves further
studies.

Our results in their present form are applicable only in higher
dimensions $d\geq 3$. One can easily check that the expression
$(1/\Box)V$ is not well defined in low dimensions $\left(
d\leq2\right)$. Note, that the logarithmic kernel of the Green's
function of the massless field is defined in $d=2$ only up to an
additive constant. Moreover, the convolution of the Green's
function with the potential makes sense in $d=2$ only if the
latter is the total derivative of some other function
\cite{CPTII}, as, for instance, the 2-dimensional curvature scalar
in the Polyakov action \cite{Polyakov}. Thus, the extension of our
results to low-dimensional models, where other calculational
schemes completely fail, is especially important. This will be
done in the forthcoming paper \cite{infrII}.

Another possibility is the generalization of our technique to
potentials with isolated zeroes in the interior of spacetime. Even
more interesting is the situation when the potential becomes
negative $V(x)<0$ in some spacetime domains. In this case there is
a tachionic instability, and it is worth deriving quantitative
criteria describing this instability in terms of the properties of
$V(x)$.

Finally, it is important to generalize our results to curved
spacetimes and in addition consider fields with non-trivial
spin-tensor structure. All these issues are addressed in
\cite{infrII}. The nonlocal effective action can then be applied
to study interesting physical problems, like quantum black holes
evaporation \cite{MWZ}, quantum cosmology, etc.

It is worth mentioning that the developed technique for late time
asymptotics of the heat kernel could also be useful in statistical
physics for calculating low temperature partition functions.

\setcounter{section}{0}
\renewcommand{\theequation}{\Alph{section}.\arabic{equation}}
\renewcommand{\thesection}{\Alph{section}.}

\section{Integral equation for $W$}

In this Appendix we derive the integral form of the equation
    \begin{equation}
    \frac{\partial W(s|\,x,y)}{\partial s}
    +\frac{(x-y)^{\mu }}{s}
    \nabla _{\mu}W(s|\,x,y)
    -\Box W(s|\,x,y)=f(s|\,x,y),  \label{A1}
    \end{equation}
where
    \begin{equation}
    f(s|\,x,y)\equiv V(x)
    -(\nabla W(s|\,x,y))^{2},  \label{A2}
    \end{equation}
and $W(s=0|\,x,y)=0$. With this purpose we first introduce the new
function $\tilde{W}$
    \begin{equation}
    W(s|x,y)=e^{-s\Box }\tilde{W}(s|\,x,y).  \label{A3}
    \end{equation}

Using the relation
    \begin{equation}
    e^{s\Box }(x-y)^{\mu }e^{-s\Box }
    =(x-y)^{\mu }+2s\nabla ^{\mu }  \label{A4}
    \end{equation}
one finds that this function satisfies the following equation
    \begin{equation}
    \frac{\partial \tilde{W}}{\partial s}
    +\frac{(x-y)^{\mu }}{s}\nabla _{\mu }
    \tilde{W}=e^{s\Box }f,              \label{A6}
    \end{equation}
which does not contain anymore $\Box $-term on the left hand side.
To write down the formal solution of this equation in terms of the
''source'' term $f=f(s|\,x,y)$, let us introduce the
characteristic curve $\bar{x}^{\mu }(t)$ of (\ref{A6}), which
satisfies the equation
    \begin{equation}
    \frac{d\bar{x}^{\mu }(t)}{dt}
    =\frac{(\bar{x}(t)-y)^{\mu }}{t},  \label{A7}
    \end{equation}
with the boundary conditions
    \begin{equation}
    \bar{x}^{\mu }(t=0)=y^{\mu },\,\,\,
    \bar{x}^{\mu }(t=s)=x^{\mu }.          \label{A8}
    \end{equation}
The solution of (\ref{A7}) is
    \begin{equation}
    \bar{x}^{\mu }(t)=y^{\mu }
    +\frac{(x-y)^{\mu }}{s}t.      \label{A9}
    \end{equation}
The total derivative of $\tilde{W}(t|\bar{x}(t),y)$ with respect
to $t$ along this characteristic curve is then equal to
    \begin{equation}
    \frac{d}{dt}\tilde{W}(t|\bar{x}(t),y)
    =\left[ \,\frac{\partial }{\partial t}+
    \frac{(x-y)^{\mu }}{t}
    \frac{\partial }{\partial \bar{x}^{\mu }}\right]
    \tilde{W}(t|\bar{x},y)
    =e^{t\overset{-}{\Box }}f(t|\,\bar{x}(t),y),     \label{A10}
    \end{equation}
where $\overset{-}{\Box }\equiv \partial ^{2}/\partial
\bar{x}^{\mu }\partial \bar{x}_{\mu }$. Integrating this equation
from $0$ to $s$ with the initial condition $\tilde{W}=0$ at $t=0$
and taking into account the boundary conditions (\ref{A8}) for
$\bar{x}(t)$, one gets
    \begin{equation}
    \tilde{W}(s|\,x,y)=\int_{0}^{s}dt\,
    e^{t\overset{-}{\Box }}\,
    f(t|\,\bar{x}\left( t\right) ,y).  \label{A11}
    \end{equation}
Returning to the original $W$ which is related to $\tilde{W}$ via
(\ref{A3}) and taking into account that
$\Box=\left(t/s\right)^{2}\overset{-}{\Box}$ we finally obtain:
    \begin{equation}
    W(s|\,x,y)=s\int_{0}^{1}d\alpha \,
    e^{s\alpha (1-\alpha )\overset{-}{\Box}}
    f(s\alpha |\,\bar{x},y)\,
    \Big|_{\,\bar{x}=\alpha x+(1-\alpha )y},  \label{A12}
    \end{equation}
where instead of $t$ the new integration variable $\alpha =t/s$
was introduced. This is exactly the integral form (\ref{intform})
of the equation (\ref{A1}) that we used in Sect.3.

\section{Covariant perturbation theory and late time behavior of the
functional trace of the heat kernel}

Here we consider the nonlocal covariant perturbation theory (CPT)
of \cite{CPTI,CPTII,CPTIII}. In CPT the functional trace of the
heat kernel for the covariant second order differential operator
is expanded as nonlocal series in powers of the potential $V$ with
explicitly calculable coefficients - nonlocal form factors
$F_{n}(s|\,x_{1},x_{2},...x_{n})$. Their leading asymptotic
behavior at large $s$ was obtained in \cite{CPTII}. Here we
calculate  them up to the first subleading order in $1/s$
inclusive for a simple operator $\widehat{F}=\Box-V$. Then we
explicitly perform infinite summation of power series in the
potential to obtain the nonlocal and nolinear expression
(\ref{sol}) for late time behavior of ${\rm Tr}\,K(s)$.

According to \cite{CPTII} the heat kernel trace is local in the first two
orders of the perturbation theory in the potential (\ref{CPT1})
    \begin{eqnarray}
    &&{\rm Tr}\,K_{0}(s)=\frac{1}{(4\pi s)^{d/2}}\int dx, \\
    &&{\rm Tr}\,K_{1}(s)=-\frac{s}{(4\pi s)^{d/2}}\int dx\,V(x),
    \end{eqnarray}
and in higher orders it reads as
    \begin{equation}
    {\rm Tr}\,K_{n}(s)=\frac{(-s)^{n}}{(4\pi s)^{d/2}n}
    \int dx\,\big<e^{s\Omega_{n}}\big>\,
    V(x_{1})V(x_{2})...V(x_{n})
    \Big|_{x_{1}=...=x_{n}=x},\,\,\,n\geq 2.  \label{B1}
    \end{equation}
Here $\Omega _{n}$ is a differential operator acting on the product of $n$
potentials
    \begin{equation}
    \Omega _{n}=\sum\limits_{i=1}^{n-1}
    \nabla_{i+1}^{2}+2\sum\limits_{i=2}^{n-1}
    \sum\limits_{k=1}^{i-1}\beta _{i}(1-\beta_{k})
    \nabla _{i+1}\nabla _{k+1},                    \label{B2}
    \end{equation}
expressed in terms of the partial derivatives labelled by the
indices $i$ implying that $\nabla _{i}$ acts on $V(x_{i})$. It is
assumed in (\ref{B1}) that after the action of all derivatives on
the respective terms all $x_{i}$ are set equal to $x$. It is also
assumed that the spacetime indices of all derivatives $\nabla
=\nabla ^{\mu }$ are contracted in their bilinear combinations,
$\nabla _{i}\nabla _{k}\equiv \nabla _{i}^{\mu }\nabla _{\mu\,k}$.
The differential operator (\ref{B2}) depends on the parameters
$\beta _{i}$, $i=1,...n-1$, which are defined in terms of the
parameters $\alpha_{i}$, $i=1,...n$ as
    \[
    \beta _{i}=\alpha _{i+1}+\alpha _{i+2}+...+\alpha _{n},
    \]
and the angular brackets in $\big<e^{s\Omega _{n}}\big>$ imply
that this operator exponent is integrated over compact domain in
the space of $\alpha $-parameters
    \[
    \big<e^{s\Omega _{n}}\big>\equiv
    \int\limits_{\alpha _{i}\geq 0}d^{n}\alpha
    \,\delta \Big(\sum_{i=1}^{n}\alpha _{i}-1\Big)\,
    \exp (s\Omega _{n}).
    \]

The late time behavior of ${\rm Tr}\,K_{n}(s)$ is thus determined
by the asymptotic behavior of this integral at
$s\rightarrow\infty$, which can be calculated using the Laplace
method. To apply this method, let us note that $\Omega _{n}$ is a
{\em negative} semidefinite operator (this is shown in the
Appendix B of \cite{CPTII}) which degenerates to zero at $n$
points of the integration domain: $(0,...0,\alpha _{i}=1,0,...0)$,
$i=1,...n$. Therefore the asymptotic expansion of this integral is
given by the contribution of the corresponding $n$ maxima of the
integrand at these points. The integration by parts in (\ref{B1})
is justified by the formal identity
$\nabla_{1}+\nabla_{2}+...+\nabla _{n}=0$ . Using it one can show
that the contributions of all these maxima are equal, so that it
is sufficient to calculate only the contribution of the point
$\alpha _{1}=1$, $\alpha _{i}=0$, $i=2,...n$. In the vicinity of
this point it is convenient to rewrite the expression for
$\Omega_{n}$ in terms of the independent $(n-1)$ variables
$\alpha_{2},\alpha _{2},...\alpha _{n}$, the remaining
$\alpha_{1}=1-\sum_{i=2}^{n}\alpha _{i}$,
    \begin{equation}
    \Omega _{n}=\sum_{i=2}^{n}\alpha _{i}D_{i}^{2}
    -\sum_{m,k=2}^{n}\alpha_{m}\alpha _{k}D_{m}D_{k},
    \end{equation}
where the operator $D_{m}$ is defined as
    \begin{equation}
    D_{m}=\nabla _{2}+\nabla _{3}+...+\nabla _{m},
    \,\,\,m=2,...n.                              \label{D}
    \end{equation}

Substituting this expression for $\Omega _{n}$ in (\ref{B1}) and
expanding in powers of the term bilinear in $\alpha$-parameters
one gets
    \begin{eqnarray}
    &&{\rm Tr}\,K_{n}(s)=
    \frac{(-s)^{n}}{(4\pi s)^{d/2}}\int
    dx\,\int_{0}^{\infty }d^{n-1}\alpha \,
    \exp \Big(s\sum_{i=2}^{n}\alpha_{i}D_{i}^{2}\Big)  \nonumber \\
    &&\qquad \qquad \qquad \qquad \qquad
    \times \left( 1-s\sum_{m,k=2}^{n}\alpha_{m}
    \alpha _{k}D_{m}D_{k}+...\right) \,V_{1}V_{2}...V_{n}.
    \end{eqnarray}
Here $1/n$ factor disappeared due to the contribution of $n$ equal
terms and the range of integration over
$\alpha_{2},...\alpha_{n}$, $\sum_{i=2}^{n}\alpha _{i}\leq 1$, was
extended to all positive values of $\alpha _{i}$. This is
justified since the error we make by extending the integration
range is exponentially small and goes beyond the the accuracy of
asymptotic expansion in inverse powers of $s$. The second term in
the round brackets can be rewritten in terms of the derivatives
with respect to $D_{m}^{2}$ acting on the exponential, so that
    \begin{eqnarray}
    &&{\rm Tr}\,K_{n}(s)=\frac{(-s)^{n}}{(4\pi s)^{d/2}}
    \int dx\,\left( 1-\frac{1}{s}\sum_{m,k=2}^{n}D_{m}D_{k}
    \frac{\partial }{\partial D_{m}^{2}}\frac{
    \partial }{\partial D_{k}^{2}}+...\right)   \nonumber \\
    &&\qquad \qquad \qquad \qquad \qquad \times
    \int_{0}^{\infty }d^{n-1}\alpha
    \,\exp \Big(s\sum_{i=2}^{n}
    \alpha _{i}D_{i}^{2}\Big)\,V_{1}V_{2}...V_{n}.
    \end{eqnarray}
In this form it is obvious that further terms of expansion in
powers of the quadratic in $\alpha $ part of $\Omega _{n}$ bring
higher order corrections of the $1/s$-series. Doing the integral
over $\alpha$ here and performing differentiations one obtains
    \begin{eqnarray}
    &&{\rm Tr}\,K_{n}(s)=\frac{1}{(4\pi s)^{d/2}}\int dx\,
    \left[ -s\frac{1}{D_{2}^{2}...D_{n}^{2}}
    +2\sum_{m=2}^{n}\frac{1}{D_{2}^{2}...D_{m-1}^{2}}\,
    \frac{1}{(D_{m}^{2})^{2}}\,
    \frac{1}{D_{m+1}^{2}...D_{n}^{2}}\right. \nonumber \\
    &&\qquad +2\sum_{m=2}^{n-1}\sum_{k=m+1}^{n}
    \frac{1}{D_{2}^{2}...D_{m-1}^{2}}
    \,\frac{D_{m}^{\mu }}{(D_{m}^{2})^{2}}\,
    \frac{1}{D_{m+1}^{2}...D_{k-1}^{2}}\,
    \frac{D_{k\mu }}{(D_{k}^{2})^{2}}\,
    \frac{1}{D_{k+1}^{2}...D_{n}^{2}}     \nonumber \\
    &&\qquad \qquad \qquad \qquad \qquad \qquad
    \left. +O\left( \frac{1}{s}
    \right) \right] \,V_{1}V_{2}...V_{n}.  \label{bigequation}
    \end{eqnarray}

The first term in the square brackets gives the leading order term
of the late time expansion. It can be further transformed by
taking into account that any operator $D_{m}$ defined by (\ref{D})
acts as a partial derivative only on the group of factors
$V_{m}V_{m+1}...V_{n}$ in the full product $V_{1}...V_{n}$,
$D_{m}V_{1}...V_{n}=V_{1}...V_{m-1}\nabla (V_{m+1}...V_{n})$.
Therefore all the operators understood as {\em acting to the
right} can be ordered in such a way
    \[
    {\rm Tr}\,K_{n}(s)=-\frac{s}{(4\pi s)^{d/2}}
    \int dx\,V_{1}\frac{1}{D_{n}^{2}}
    V_{n}\frac{1}{D_{n-1}^{2}}V_{n-1}...
    \frac{1}{D_{2}^{2}}V_{2}+O\left( \frac{1}{s^{d/2}}\right)
    \]
that the labels of $D_{m}^{2}$'s can be omitted and all $D_{m}^{2}$ can be
identified with boxes also acting to the right
    \begin{equation}
    {\rm Tr}\,K_{n}(s)=-\frac{s}{(4\pi s)^{d/2}}
    \int dx\,\underbrace{V\frac{1}{\Box }V\frac{1}{\Box }...
    V\frac{1}{\Box }}_{n-1}\,V(x)
    +O\left( \frac{1}{s^{d/2}}\right) .  \label{b1}
    \end{equation}
Infinite summation of this series is not difficult to perform
because this is the geometric progression in powers of the
nonlocal operator $V(1/\Box )$ and
    \[
    {\rm Tr}\,K(s)={\rm Tr}\,K_{0}(s)
    -\frac{s}{(4\pi s)^{d/2}}\int
    dx\,\sum_{n=0}^{\infty }
    \left( V\frac{1}{\Box }\right) ^{n}V(x)
    +O\left(\frac{1}{s^{d/2}}\right),
    \]
or
    \begin{equation}
    {\rm Tr}\,K(s)=\frac{1}{(4\pi s)^{d/2}}
    \int dx\,\Big(1-s\Box \frac{1}{\Box-V}\,V(x)
    +O(s^{0})\Big).                               \label{b2}
    \end{equation}
The second term here looks as a total derivative. However, it does
not vanish because this is a derivative of the nonlocal expression
and the corresponding surface term does not vanish at infinity in
view of the Green's function asymptotics. This term can be
rewritten as
    \begin{equation}
    \Box \frac{1}{\Box -V}\,V(x)
    =V(x)+V\frac{1}{\Box -V}\,V(x)=V\Phi (x),     \label{b3}
    \end{equation}
and, therefore,
    \begin{equation}
    {\rm Tr}\,K(s)=\frac{1}{(4\pi s)^{d/2}}
    \int dx\,\Big(1-s\,V\Phi (x)+O(s^{0})\Big),  \label{b4}
    \end{equation}
where $O(s^{0})$ denotes the subleading in $s$ terms which depend
on the potential in a nontrivial way. They are given by infinite
resummation over $n$ of the second and third terms in square
brackets of Eq.(\ref{bigequation}). Remarkably, this summation can
again be explicitly done. In this case one has to sum multiple
geometric progressions.

Indeed, the second term of (\ref{bigequation}) gives rise to the series
    \begin{equation}
    \frac{2}{(4\pi s)^{d/2}}\int dx\,
    \sum_{n=2}^{\infty }\sum_{m=2}^{n}V
    \underbrace{\frac{1}{\Box }V...
    \frac{1}{\Box }V}_{n-m}\frac{1}{\Box ^{2}}
    \underbrace{V\frac{1}{\Box }...
    V\frac{1}{\Box }}_{m-2}\,V(x).                  \label{1}
    \end{equation}
By summing the two geometric progressions with respect to
independent summation indices $0\leq n-m<\infty $ and $0\leq
m-2<\infty$ one finds that this series reduces to
    \begin{equation}
    \frac{2}{(4\pi s)^{d/2}}\int dx\,V\,
    \frac{1}{(\Box -V)^{2}}\,V(x),                    \label{2}
    \end{equation}
which after the integration by parts amounts to
    \begin{equation}
    \frac{2}{(4\pi s)^{d/2}}\int dx\,
    \left( \frac{1}{\Box -V}\,V(x)\right) ^{2}=
    \frac{2}{(4\pi s)^{d/2}}\int dx\,
    \Big(1-\Phi (x)\Big)^{2}.                  \label{B1000}
    \end{equation}

Similarly, the third term of (\ref{bigequation}) gives rise to the
triplicate geometric progression which after summation and
integration by parts reduces to
    \begin{eqnarray}
    &&\frac{2}{(4\pi s)^{d/2}}\int dx\,
    V\sum_{i=0}^{\infty }\Big(\frac{1}{\Box }V
    \Big)^{i}\frac{1}{\Box }\nabla ^{\mu }
    \frac{1}{\Box }\,\sum_{j=0}^{\infty }
    \Big(V\frac{1}{\Box }\Big)^{j}
    \frac{1}{\Box }\nabla ^{\mu }\frac{1}{\Box}
    \sum_{l=0}^{\infty }
    \Big(V\frac{1}{\Box }\Big)^{l}\,V(x)  \nonumber \\
    &&\qquad \qquad \qquad \qquad
    =-\frac{2}{(4\pi s)^{d/2}}\int dx\,
    \Big(\nabla_{\mu }\Phi (x)\Big)
    \frac{1}{\Box -V}V\frac{1}{\Box }
    \nabla ^{\mu }\Phi (x).                   \label{B100}
    \end{eqnarray}
Taking here into account that
    \begin{equation}
    \frac{1}{\Box -V}V\frac{1}{\Box }
    =\frac{1}{\Box -V}-\frac{1}{\Box }
    \end{equation}
one finds that the sum of (\ref{B1000}) and (\ref{B100}) is equal to
    \begin{eqnarray}
    &&\frac{2}{(4\pi s)^{d/2}}\int dx\,
    \left( (1-\Phi )^{2}-\nabla _{\mu }\Phi
    \frac{1}{\Box -V}\nabla ^{\mu }\Phi
    +\nabla _{\mu }\Phi \frac{1}{\Box }
    \nabla ^{\mu }\Phi \right)   \nonumber \\
    &&\qquad \qquad \qquad \qquad \qquad \qquad
    =-\frac{2}{(4\pi s)^{d/2}}\int
    dx\,\nabla _{\mu }\Phi
    \frac{1}{\Box -V}\nabla ^{\mu }\Phi ,  \label{B10000}
    \end{eqnarray}
where the cancellation of the first and the third terms takes
place after rewriting $\nabla _{\mu }\Phi $ in the third term as
$\nabla _{\mu }(\Phi -1)$ and integrating it by
parts\footnote{Straightforward integration by parts of
$\nabla_{\mu }\Phi (1/\Box )\nabla^{\mu }\Phi $ is impossible
because $\Phi (x)$ does not vanish at $|x|\rightarrow \infty$,
while $\Phi (x)-1$ does.}. Together with (\ref{b4}) the
contribution (\ref{B10000}) forms the nonlinear and nonlocal late
time expression for the heat kernel trace (\ref{sol}) up to the
first subleading order in $1/s$ inclusive.

\section*{Acknowledgement}

A.O.B. is grateful for hospitality of the Physics Department of
LMU, Munich, where a major part of this work has been done. The
work of A.O.B. was also supported by the Russian Foundation for
Basic Research under the grant No 02-02-17054. V.M. thanks SFB375
for support.

\end{document}